# Examining the Effects of Objective Hurricane Risks and Community Resilience on Risk Perceptions of Hurricanes at the County Level in the U.S. Gulf Coast: An Innovative Approach

Wanyun Shao

Auburn University at Montgomery; wshao@aum.edu

Maaz Gardezi

South Dakota State University; Maaz.Gardezi@sdstate.edu

Siyuan Xian

Princeton University; sxian@princeton.edu

# ABSTRACT

Communities' risk perceptions can influence their abilities to cope with coastal hazards such as hurricanes and coastal flooding. Our study presents an initial effort to examine the relationship between community resilience and risk perception at the county level, through innovative construction of aggregate variables. Utilizing the 2012 Gulf Coast Climate Change Survey merged with historical hurricane data and community resilience indicators, we first apply a spatial statistical model to construct a county-level risk perception indicator based on survey responses. Next, we employ regression to reveal the relationship between contextual hurricane risk factors and community resilience, on one hand, and county-level perceptions of hurricane risks, on the other. Results of this study are directly applicable in the policy-making domain as many hazard mitigation plans and adaptation policies are designed and implemented at the county level. Specifically, two major findings stand out. First, the contextual hurricane risks represented by peak height of storm surge associated with the last hurricane landfall and land area exposed to historical storm surge flooding positively affect county-level risk perceptions. This indicates that hurricanes' another threat – wind risks – need to be clearly communicated with the public and fully incorporated into hazard mitigation plans and adaptation policies. Second, two components of community resilience – higher levels of economic resilience and community capital – are found to lead to heightened perceptions of hurricane risks, which suggests that concerted efforts are needed to raise awareness of hurricane risks among counties with less economic and community capitals.

Keywords: county-level perceptions of hurricane risks; objective hurricane risks; community resilience; U.S. Gulf Coast

## INTRODUCTION

The coupling effects of changing climate and rising concentration of population and assets in the coastal regions have increased the threat of potential damages. Among all natural hazards of the past century, hurricanes and its induced flooding are the most costly in the United States (Aerts et al. 2014; Meyer et al. 2014; Michel-Kerjan 2015). The recent two most costly natural disasters in the US are Hurricane Katrina (2005) and Super Storm Sandy (2012) that both caused severe damages in coastal communities (Hatzikyriakou et al. 2015; Xian, Lin, and Hatzikyriakou 2015). The recent Hurricanes Harvey, Irma and Maria have raised more concern for this hazard. Scientific modeling projections show that the destructiveness and intensity of hurricanes may become stronger over time (Emanuel 2005; Emanuel 2013). The hurricane-induced coastal floods are also predicted to become more frequent and intense in the future (Lin and Emanuel 2015). Thus, there is an urgent need for coastal communities to prepare well for increasing hurricane and its related flooding risks through mitigation and adaptation measures (Xian, Lin, and Kunreuther 2017).

Recent empirical research has found that peoples' motivation of voluntary risk mitigation and adaptation is low unless actual risk can be perceived (Botzen et al. 2009; Kunreuther and Slovic 1978; Kunreuther and Weber 2014; Lindell and Hwang 2008; Shao et al. 2017b; Shao et al. 2017c). Risk perceptions can influence peoples' willingness to respond to, recover from, and adapt to hurricanes and other extreme events (Hertwig et al. 2004; Kasperson et al. 1988; Loewenstein et al. 2001). For instance, individuals' perceptions of flood risks have direct effects on adaptive measures, such as hazard mitigation (Huang et al. 2012) and evacuation behaviors (Ge, Peacock and Lindell 2011) and voluntary purchase of flood insurance (Shao et al. 2017b), and policy support for incentives for relocation and funding for education programs on emergency planning and evacuation (Shao et al. 2017c). Therefore, examining perceptions towards hurricane-related risks can provide policy makers with insights into mitigation/adaptation policy support and public willingness to take follow-up actions. Risk perceptions are socially constructed and can be influenced by various factors such as past experiences of natural hazards (Shao 2016; Shao and Goidel 2016; Shao et al. 2017a), social relations, emotional reactions to risky situations, group cultural values, and community ways of life (Douglas and Wildavsky 1983; Hertwig et al. 2004; Weber and Stern 2011; Loewenstein et al 2013; Weber and Hsee 1998).

While extensive research on individual risk perceptions and their relationships with personal actions has been conducted in the context of hurricane-induced coastal floods (Lindell and Hwang, 2008; Shao et al. 2017a; Shao et al. 2017b; Shao et al. 2017c;), little statistically rigorous and theoretically informed investigation has focused on examining the determinants of community level hurricane-related risk perceptions among coastal counties in the U.S. Gulf Coast. The term community-level risk perceptions refer to an aggregation of individual risk perceptions, rather than collective risk perceptions shared by all individuals in a community. This aggregate measure of risk perceptions constructed through statistical transformation partially reflects the collective perception as well as variations that exist among the population in a community. Given the diversity of perceptions in a community, it may be challenging to identify a unified collective measure of risk perceptions. Instead, this aggregate measure built from individual input may be more realistic to reflect community-level risk perceptions. The rationale of studying community-level risk perceptions is as follows: (1) public policies are designed and implemented at an aggregate unit (e.g., county, state, nation). The understanding of individual risk perceptions does not directly translate into knowledge of aggregate risk perceptions. In other words, the examination of community risk perceptions can provide policy makers with insights that can be directly applied, and (2) aggregate characteristics such as demographics and socio-economic composition are the integral components of a policy-making system. Given the strong link between risk perceptions and actions to cope with risks, aggregate risk perceptions should also be incorporated into policy-making process. In order for various social and cognitive components to be smoothly integrated into one system, they must be conceptualized and measured at the same unit. It is thus necessary to build an aggregate indicator that represents community-level risk perceptions.

Meanwhile, a substantial amount of empirical evidence indicates the significant impact of the objective environment on individual environmental risk perceptions (Botzen et al. 2009; Shao et al. 2014; Shao 2016; Shao et al. 2017a). It can be equally important to investigate how the objective environment shapes risk perceptions on the aggregate level. In the present study, we focus on understanding aggregate hurricane-related risk perceptions. The objective environment consists of social and natural contexts. Community resilience can largely represent a social context within which risk perceptions may be formed and influenced. Community resilience include social resilience, economic resilience, infrastructure resilience, community capital resilience, institutional resilience, and environmental resilience. The natural context in this study is represented by hurricane-related risks including: land area exposed to hurricane wind risks, land area exposed to storm surge, maximum wind speed and peak height of storm surge associated with the last hurricane landfall.

In this article, we fill the existing vacancy in the literature by examining how community resilience and hurricane-related risks can influence community-level risk perceptions. In particular, we study the effects of various hurricane-related risk factors and community resilience on the county-level hurricane-related risk perceptions among coastal counties in the U.S. Gulf Coast. It should be noted that the community is represented by county in this study. Our choice of county as the focal geographic unit is based on the following considerations: (1) the areas of coastal counties along the U.S. Gulf Coast are relatively homogeneous in comparison to that of zip codes; (2) institutional processes including many policies and decisions are made at the county level, such as disaster relief and thus results at this aggregate level can be directly applicable in the policy-making domain, and (3) many existing community-level socio-economic-institutional indicators such as social vulnerability index (Cutter 2003) and community resilience index (Cutter, et al. 2011) are constructed at the county level. Thus, it would be of interest and convenience to build our community-level risk perceptions on the same geographic level for direct examinations of effects of these comprehensive objective community characteristics on risk perceptions.

Our study highlights the relationship between county-level risk perception and community resilience at the county level with the potential goal of revealing the dynamic socio-cognitive processes that are triggered by building or eroding community resilience in future studies. This article is organized as follows: the conceptual framework is first presented, with each component being discussed. The data and methods are presented in the following section. Results of the analyses are discussed subsequently. This article concludes with a summary of findings, discussions of implications, and a path forward for future studies.

# Conceptual Framework: Risk Perceptions, Contextual Risks, and Community Resilience

Our conceptual framework consists of three major parts: contextual risks, community resilience, and aggregate risk perceptions of hurricanes. Figure 1. lays out the relationships among these three components. These hypothesized relationships are based on the past literature and explored in this study.

**Figure 1. Conceptual Framework**

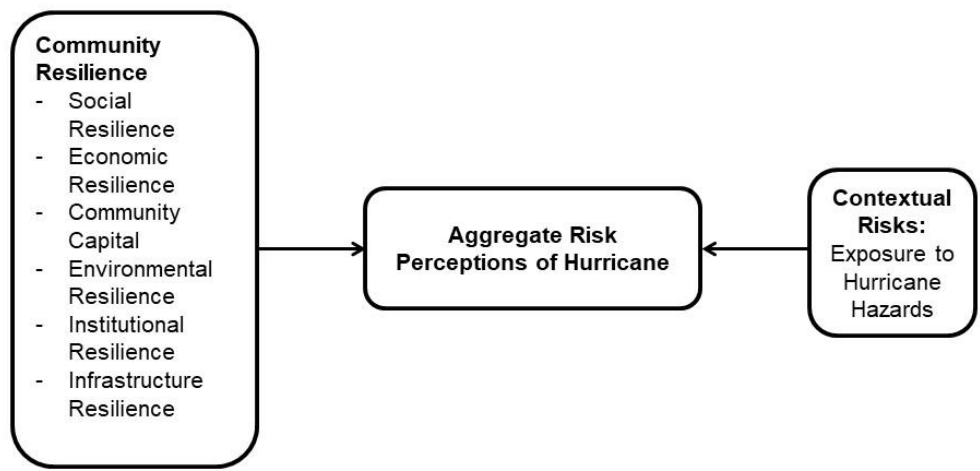

## Contextual risks and risk perceptions

Lindell and Perry (2012) lay out a theoretical framework in which the environmental context constitutes the initial stage of a decision process. The environmental setting provides cues that can trigger perceptions of environmental threats. Numerous empirical studies have found evidence to support such a link between contextual risks and risk perceptions on the individual level. For instance, increasing summer and winter temperatures can lead to heightened risk perceptions of global warming (Howe, et al. 2013; Shao et al. 2014; Shao 2016). Maximum wind speed and peak storm surge height associated with the last hurricane landfall can shape coastal residents' perceptions of changing hurricane strength (Shao et al. 2017a). Hazardous fuel conditions on or near landowners' parcels can heighten their perceptions of wildfire risks (Fischer et al. 2014). In their model on private proactive adaptation to climate change, Grothmann and Patt (2005) acknowledge the significant role of contextual risks in determining risk perceptions. In this study, we aim to examine if this effect of the contextual risks can also influence aggregated perceived risks. We hypothesize that *the contextual risks represented by exposure to past hurricane-related hazards can have positive impacts on county-level perceptions of hurricane risks.*

**Community resilience and risk perception**

Of interest to this study is to examine how community resilience can influence community-level risk perceptions. Cutter et al. (2008) defined community resilience as "continual learning and taking responsibility for making better decisions to improve the capacity to handle hazards." Therefore, a resilient community is constantly adapting to changing conditions through continuous learning, flexibility, and risk management. These attributes are indicative of peoples' adaptive capacities, i.e. their ability to respond proactively to dynamic future and uncertain events and environments (Eakin et al. 2016). Although considerable research has been devoted to measuring resilience through changes in adaptive capacity (Engle and Lemos 2010), rather less attention has been paid to understand how peoples' subjective understanding of risk, such as risk perception can be incorporated into community resilience. The definition given by Cutter et al. (2008) implies that cognitive components such as "learning" and "decisions" should be integrated into resilience. "Continual learning" requires one to constantly seek accurate information from the external sources. "Better decisions" need to incorporate and reflect the accurate information. Between seeking external information and reflecting information lies perception. In their socio-cognitive model of private proactive adaptation to climate change, Grothmann and Patt (2005) emphasized the role of individual cognition including both risk appraisal and adaptation appraisal in enabling or impeding adaptation. Risk appraisal can result in risk perception which can determine whether the other major feature - adaptation appraisal should proceed (Grothmann and Patt 2005). Only when a threshold of risk appraisal is exceeded will adaptation appraisal be engendered (Grothmann and Patt 2005). Risk perception is thus a precondition of adaptation perception and subsequent decision and behavior. By understanding community's perception of risk and their proximity to coping thresholds, scholars can gain an indication of the magnitude of disturbance the system can withstand before its function and structure begin to lose originality (Marshall and Marshall 2007).

In this article, we argue that community-level risk perception is one of the defining system attribute in community resilience. Thus, we contend that community-level risk perception should be an integral component of community resilience. Before fully integrating risk perception into community resilience, there is a need to examine the relationship between each constituent of community resilience and community-level risk perceptions. In order to guide appropriate incorporation of risk perceptions into the construction of community resilience, it is of interest to reveal which dimensions of community resilience is/are more correlated with risk perception than others. Built on previous work, Cutter, Ash, and Emrich (2014) refined the community resilience index by including a more comprehensive suite of variables and a larger study area. Their community resilience metrics include: social resilience, economic resilience, community capital, institutional resilience, infrastructural resilience, and environmental resilience. This resilience index measures objective resilience using secondary data collected by government agencies. Given the broad scope of their community resilience index, we directly use these objective indicators to represent the societal and institutional contexts within which risk perceptions are formed and influenced. In doing so, we attempt to reveal the impact of objective community resilience on community risk perception, a potential component of a more comprehensive resilience index. This community resilience index is comprehensive and diverse in scope. We hypothesize that *this group of indicators plays a significant role in determining perceptions of hurricane-related risks.*

In particular, social resilience, economic resilience, community capital, and environmental resilience tend to represent the innate and relatively unmalleable composites and characteristics of a community, which could require long period of time to change. Specifically, both social resilience and community capital are related to socio-demographic properties of a community's population that reflects its innate ability and conscience to cope with natural disasters. Economic resilience is meant to capture the general profile of a community's economic diversity, equality, and ties with business in other communities.

Institutional resilience and infrastructure resilience, on the other hand, tend to represent a community's capacity to make immediate adjustments to mitigate natural disasters. For instance, the 10 institutional indicators considered by Cutter, Ash, and Emrich (2014, 68) are "meant to capture aspects related to programs, policies, and governance of disaster resilience." The infrastructural indicators meanwhile capture "a myriad of physical capacities within a county" to provide emergency aid to members of a community when a disaster occurs (Cutter, Ash, and Emrich 2014). Cutter, Ash, and Emrich (2014) also suggest that compared to environmental resilience meant to represent the qualities of the environment to absorb long-term impacts engendered by slower onset disasters, infrastructural resilience can be critical for coping with shorter term disaster. While the occurrence of extreme events such as hurricanes could generate a great amount of interest among disaster management authorities and lead to immediate responses at the institutional and infrastructure level (Xian et al. 2018; Næss et al. 2005), changes in social and economic resilience often occur over longer ranging times. Our contention therefore is that *the contextual risk has substantial effects on institutional and infrastructure resilience.*

It is worth noting that the dimensions of the contextual risk are not nearly commensurate with the components of community resilience. By no means do we intend to make the contextual risk commensurate with community resilience in the present study out of the following considerations: 1). our focus is hurricane risk perceptions, not general environmental risk perceptions. To correspond to the particularity of this, risk perceptions of hurricanes should be understood within the context of a specific risk; 2). The social context as represented by community resilience here does not vary with the change of the type of natural hazards. For instance, whether a community in California is coping with wildfires or earthquakes, the social context remains.

## DATA AND METHODS

### Data

The primary data come from the 2012 Gulf Coast Climate Change Survey (Goidel et al. 2012). The survey employed stratified random sampling to draw independent samples of coastal residents across five Gulf Coast States (Florida, Alabama, Mississippi, Louisiana, and Texas). The sampling approach was designed to select residents situated in communities contending with a changing climate, such as shifts in hurricanes, droughts, flooding, as well as coastal wetland loss and erosion (Goidel et al. 2012). Landline telephone was used to collect responses from 3,856 residents. The survey measured residents' climate change-related risk perceptions, perception of government risk-reduction policies and programs, personal willingness to take actions to adapt to climate change impacts, and socio-demographic features. This survey is to date the most comprehensive survey assessment of coastal residents' perceptions of local climate shifts in the Gulf Coast region. The survey data provide county FIPS as geographic identification codes for each respondent. Community resilience come from the 2010 Baseline Resilience Indicators for Communities (BRIC) (Cutter, Ash, and Emrich 2014). Table 1 provides explanation of each community resilience component. Using the geographic information (county), we merged the community resilience data with the survey data.

**Table 1. Explanation for Each Community Resilience Indicator (Source: Cutter et al. 2014, 68)**

Social resilience is "intended to capture demographic qualities of a community's population that tend to associate with physical and mental wellness leading to increased comprehension, communication, and mobility."

Economic resilience is "intended to represent community economic vitality, diversity, and equality in compensation."

Community capital "conceptually represent the level of community engagement and involvement in local organizations and the potential for local ties and social networks that can be critical for survival and recovery during disaster."

Environmental resilience "conceptually relates to qualities of the environment that enhance absorptive capacity of coastal surges and freshwater flooding in particular."

Institutional resilience is "meant to capture aspects related to programs, policies, and governance of disaster resilience."

Infrastructure resilience estimates "the quality of housing construction and a myriad of physical capacities within a county to house the displaced, provide emergency medical care, facilitate evacuations, and maintain schooling activities, among other disaster-relevant infrastructural capacities."

The environmental data come from four sources. The hurricane track and wind speed data come from the National Hurricane Center (HURDAT best track data). The high water mark data is extracted from the Program for the Study of Developed Shorelines (PSDS) Storm Surge Database at Western Carolina University. Data on maximum wind speed was downloaded from the ICAT Damage Estimator Database. Lastly, data on storm surge events were extracted from a comprehensive database developed by SURGEDAT. With the geographic information for each respondent, we merge the environmental data with the survey data at the county level.

**Dependent variable:**
**Individual-level survey responses**

Of interest to this study is to extract information about residents' hurricane-related risk perceptions. Perceptions of hurricane-related risk factors are based on residents' responses to three survey questions about changes in number and strength of hurricanes, and amount of flooding (Table 2).

Set in the coastal region, hurricane landfalls can generate high storm surge which can lead to coastal flooding. Between 2005 and 2010 (recent past for respondents of this survey), the U.S. Gulf Coast experienced devastating effects from storm surge flooding caused by Hurricanes Katrina (2005), Rita (2005), and Ike (2008). Given the close correlation between wind force and induced coastal flooding, it is reasonable to assume that perceptions of flooding amount are related to perception of hurricane risk.

**Table 2. Description of survey items used in the study for constructing individual-level risk perception variable (n=3852)**

| Statement | Decreased | Unchanged | Increased | Don't Know (NA) |
|---|---|---|---|---|
| Would you say that the number of hurricanes that have impacted your local community have increased, decreased, or stayed about the same as in the past? | 23.36% | 52.74% | 21.23% | 2.64% |
| Would you say that the hurricanes that do impact your local community are stronger, not as strong or about as strong as hurricanes in the past? | 13.43% | 42.37% | 36.01% | 8.19% |
| Would you say flooding in your local community has increased, decreased, or stayed about the same? | 17.06% | 56.92% | 23.40% | 2.59% |

To validate this speculation, an exploratory factor analysis (EFA) with varimax rotations is used to condense information from these variables into a single item measuring hurricane-related risk perception *(risk perception)*. Formally, factor scores for residents' hurricane-related risk perceptions are estimated using the following equation:

$$F_{zr} = \Phi_{zr}\Omega \qquad \text{Equation (1)}$$

Where:

$F_{zr}$ is a matrix of standardized factor scores for risk perceptions

$\Phi_{zr}$ is a matrix of standardized observed scores for risk perceptions (z-scores)

$\Omega$ is a matrix of factor score weights

The z-scores for risk perceptions were normalized by this equation:

$$F_{zr_i} = \frac{F_{re_i} - F_{re\,min}}{F_{re\,max} - F_{re\,min}}, \qquad \text{Equation (2)}$$

Where $F_{re_i}$ denotes the standardized factor score for risk perception for resident $i$. Table 3 provides a description of factor loadings, eigenvalue, and communalities for the factor scores. As illustrated in Table 2, factor loading and communalities are in the moderate range (0.4 – 0.6) for the three survey items included in the analysis. A one factor solution is chosen for its suitability in interpreting results and for constructing a latent measure of residents' perceptions hurricane-related risks.

**Table 3. Summary of factor loadings and communalities from principal axis factor analysis for perception of flooding risk (varimax rotation, 1 factor solution)**

| Perceptions of Flooding Risk | Factor loading | $h^2$ |
|---|---|---|
| Perceptions of hurricane number | .63 | .40 |
| Perceptions of hurricane strength | .54 | .30 |
| Perceptions of flooding amount | .45 | .20 |
| Eigenvalue | 0.90 | |
| Proportion variance explained | .30 | |

Notes. $h^2$ = communality

**Statistical method for constructing the dependent variable: County-level risk perception score**

First, average, first quantile, and third quantile of risk perception scores are calculated for each county, respectively. An average entails summing risk perceptions of all residents in county *j* and then they are divided by the number of respondents in that county. One issue with using average and quantile-based risk perception scores is that these are calculated for unequal number of residents in each county. Figure 2 shows the variation in the number of respondents in each county included in the study area. It illustrates that sample size in each county varies from as low as 1 to as high as 350.

To compensate the wide variation displayed among counties, this article uses an approximation technique to calculate the county-level simple random sample variance. The approximation consists of three steps: 1) the variance among all residents is calculated; 2) the sample variance of the county mean is obtained by dividing overall variance by total number of residents in each county; 3) the first two steps are repeated for each county so that an approximate county-level simple random sample variance was calculated. Intuitively, this approximation ensures that counties with fewer survey respondents have larger sampling variances.

Moran's I is calculated to test the null hypothesis that no spatial correlation existed among county-level risk perception scores. This study uses queen neighbors with row-standardized weights to estimate the Moran's I. The equation for estimating the Moran's I (eq. 3) is given below:

$$I = \frac{n}{\sum_{i=1}^{n}\sum_{j=1}^{n} w_{ij}} \frac{\sum_{i=1}^{n}\sum_{j=1}^{n} w_{ij}(y_i - \bar{y})(y_j - \bar{y})}{\sum_{i=1}^{n}(y_i - \bar{y})} \quad \text{Equation 3}$$

Where, $y_i$ is the ith county score, $y_j$ is the jth county score, $\bar{y}$ is the overall mean of the study area and $w_{ij}$ represents the spatial weight between county i and j.

**Figure 2. The number of respondents per county in the study area**

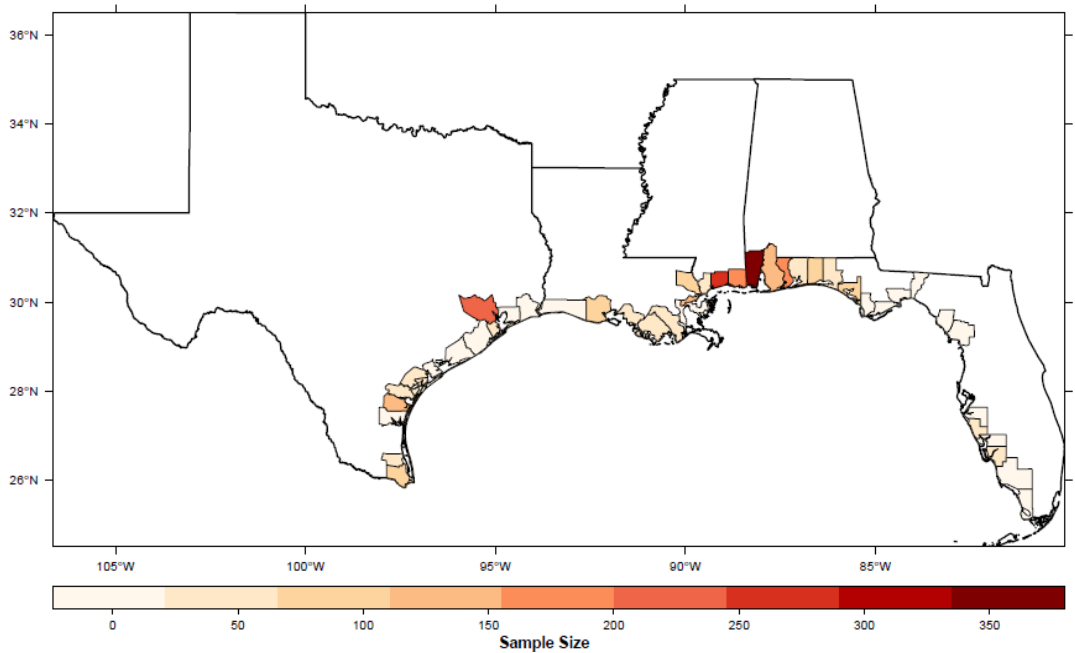

We then use a spatial Conditional Autoregressive (CAR) model to spatially smooth county-level risk perception indicators using a maximum likelihood estimation technique. First, suitable neighbors for each county are identified by specifying a queen neighbor structure, i.e. counties sharing a boundary point were taken as neighbors (Bivand, Pebesma and Gómez-Rubio 2008). A queen neighbor structure is selected for ensuring that each county was assigned at least one neighbor. This neighboring structure ensures that the prediction for a county will also include contributions from at least one spatial neighbor. On average, there are almost 2 neighboring links for every county. After establishing the set of neighbors, spatial weights are assigned to each neighbor relationship. Binary weights are assigned to ensure that the structure can define symmetry needed for estimating a CAR model. A binary weight structure assigns weight of 1 to each neighbor and 0 to non-neighbor relationship. Thus, binary weights differentiate the influence of observations—those with many neighbors are more influential (up-weighted) compared to those with few neighbors.

The Conditional Autoregressive model (CAR) is used to spatially smooth (1) average, (2) first quantile, and (3) third quantile of residents' risk perception scores for each county ($Y_i$). The CAR model assumes that the characteristics of an area—here county-level risk perceptions (average or quantiles)—is influenced by its neighbors of neighbors, etc. Formally, the CAR model can be written as:

$$Y_i | Y_{-i} = X_i \beta + \sum_{j=1}^{N} d_{ij} (Y_j - X_j \beta) + e_i \qquad \text{Equation 4}$$

Where $\beta$ is the element of the spatial dependence matrix expressing dependence among counties and $Y_{-i}$ is treated as having fixed values when specifying the distribution of $Y_i$. The variance of Y is specified as:

$$\text{Var}[Y] = (I - \Phi)^{-1} \sum v \qquad \text{Equation 5}$$

For a valid variance-covariance matrix, $\Phi$ must be symmetric, so that $d_{ij}=d_{ji}$. To fulfill the second constraint, the study uses binary weights that ensured symmetry in the neighboring structure. Based on the criteria that lowest AIC and residual variance indicate the best model fit, this study decides to use the spatially smoothed values of average risk perception as the dependent variable. The smoothed values of county-level perceptions are the sum of non-spatial and spatial fitted values, including contributions from spatial neighbors. Here, all values were smoothed to the global mean.

**Independent variables:**

**Objective Hurricane Risks**

Objective hurricane risks are represented by exposure to past hurricane-related hazards. There are two categories of measures: (1) estimates of exposed areas to storm surge-induced floods and hurricane winds, and (2) peak height of storm surge and maximum wind speed of the most recent hurricane landfall. We contend that risk perception is more affected by recent events than historical records, according to availability bias (Tversky and Kahneman 1974) and previous empirical evidence (Hertwig et al., 2004; Shao et al., 2017a).

*Land Area Exposed to Hurricane Wind Risks* According to the Saffir-Simpson wind scale, hurricanes are classified into five categories. The wind speed and intensity of damage increases from category 1 to category 5. Hurricanes reaching Category 3 and higher are considered as major hurricanes because of their potential for significant threat to life and substantial damages to property. In this study, the proportion of major hurricane wind zone per county is used to represent the land area exposed to high hurricane winds. The creation of these major hurricane wind zones involves the collection of tracks for major hurricanes from 1895-2005 that either made landfall or were passed through within 100 miles of the U.S. Gulf Coast mainland. This data came from the historical hurricane track data archives, i.e. HURDAT best track data (from National Hurricane Center) that include the latitude and longitude of points with 6-hour intervals along the hurricane track and the wind speed and central pressure associated with each points (National Oceanic and Atmospheric Administration – International Best Tracks Archive for Climate Stewardship). Previous studies reveal the complexity in modeling the hurricane wind radius and entire wind field (Chavas et al. 2015) and indicated that the average diameter of hurricane force winds is 100 miles (Willoughby 2007). Accordingly, Emrich and Cutter (2011) defined the hurricane wind impact areas as 50 miles on either side of the historic track. Although there may be some limitations due to asymmetric of hurricane structure (since the stronger winds occur on the right side of a tropical cyclone due to the contribution of the transitional wind speed), the approximation can represent the average impact extent that hurricane makes to the coastal communities. Therefore, we adopted Emrich and Cutter's approach in the present study. After mapping the hurricane tracks, a 50-mile spatial buffer zones are created along the track of the hurricane. Then, spatial intersect processes in ArcGIS are performed to compute the amount of land area within the hurricane wind impact zone for each county. The land area of hurricane wind zone divided by the total land area in the county produces the percentage of land area that had been affected by the major hurricanes in the past 100 years.

*Land Area Exposed to Storm Surge* In addition to the wind impact, hurricanes can generate high storm surge that can cause substantial damage to coastal structures. To capture this aspect of hurricane risk, percentage of land exposed to storm surge per county is created. The "High Water Level Mark Dataset" illustrates the highest water level mark (from both storm surges and storm tides) recorded from 1954 - 2012 during hurricanes along the Gulf Coast. It is first converted from Excel into shapefiles by using point geometry and then converted into raster format. The contour lines with 5-feet interval is extracted from the Digital Elevation Model (DEM) with 30-meter resolution and this is downloaded from the National Map Viewer, USGS. By using the highest water level mark (35 feet), this study extracts the land area along the Gulf Coast which has an elevation of 35 feet or below. This method has also been applied to model the potential flood inundation using a static approach (Aerts et al. 2014). Finally, the percentage of land area affected by the highest storm tide per county is computed.

*Maximum Wind Speed of the Last Hurricane Landfall* This data was extracted from the ICAT Damage Estimator Database. ICAT is an insurance company that provides catastrophic insurance coverage to businesses and homeowners in the U.S. The maximum wind speed was taken as the 1-min average at 10 meters elevation during the landfall over the affected regions. It is estimated to be the final 6-hourly magnitude prior to the landfall from the HURDAT Best Track data (Jarvinen et al. 1984; Landsea et al. 2004). The resolution of the data is in 5 knots (e.g. 50 knots, 55 knots, 60 knots…). In this study, the maximum wind speed from the last landfall since 1992 was assigned to each county. This decision is based on the presumption that communities' memories with past disasters tend to be short (Viglione et al. 2014).

*Peak height of Storm Surge of the Last Hurricane Landfall* This data came from the updated version of SURGEDAT which compiles data from 62 sources and identifies 195 storm surge events with the minimum height of 1.22 meters (Needham and Keim 2012). Among the 69 tropical cyclone events that have affected the Gulf Coast since 1992, 66 events have either storm surge or storm tide data available. In this study, we use the peak storm surge height for 10 events and peak storm tide height for the other 56 events to represent the peak height of the storm surge. We assign the peak height of storm surge from the latest hurricane landfall to each county.

**Community Resilience Indicators**

Data for community resilience came from the 2010 Baseline Resilience Indicators for Communities (BRIC) (Cutter, Ash, and Emrich 2014). The BRIC index is a composite index of 61 variables that constitute community resilience at the county level. These indicators represent six broad categories including social, economic, community capital, institutional, infrastructural, and environmental resilience. BRIC has been used to examine place-specific drivers and attributes of place-specific resilience to natural hazards (Cutter, Ash, and Emrich 2014) and to observe improvements in resilience over time (Cutter et al. 2010). Of interest to this study is to use the composites of the six resilience categories as a reference point or baseline for examining the relationship between risk and resilience at the county level. Table 4 shows the descriptive statistics of county-level dependent and independent variables in the study.

**Table 4. Description of county-level variables used in the study (n=46)**

| Components | Mean | Std. Dev | Min | Max |
|---|---|---|---|---|
| County-level Risk Perception | 0.59 | 0.08 | 0.43 | 0.77 |
| Maximum Wind Speed | 92.46 | 29.04 | 40.00 | 150.00 |
| Peak Storm Surge | 11.57 | 8.36 | 1.00 | 28.00 |
| Land Exposure to Hurricane Wind | 0.70 | 0.45 | 0.00 | 1.00 |
| Land Exposure to Storm Surge | 0.60 | 0.31 | 0.05 | 1.00 |
| Social Resilience | 0.58 | 0.11 | 0.29 | 0.79 |
| Economic Resilience | 0.49 | 0.10 | 0.23 | 0.70 |
| Infrastructure Resilience | 0.46 | 0.12 | 0.00 | 0.79 |
| Community Capital Resilience | 0.52 | 0.11 | 0.31 | 0.88 |
| Institutional Resilience | 0.58 | 0.09 | 0.40 | 0.78 |
| Environmental Resilience | 0.67 | 0.14 | 0.28 | 0.99 |

**Statistical Approach to Model Hypothesized Relationships**

This study uses a Bayesian analysis for the linear regression model. A Bayesian approach combines the prior distribution with the likelihood, i.e. statistical inference is made based on both sample (data) and prior information about population characteristics (Gelman et al. 2004). In this paper, the real benefits of using a Bayesian over a traditional or frequentist approach for linear regression are provided by use of rigorous methods of model selection and checking. Although we are using non-informative priors for our regression, this is only a placeholder, and can be updated when prior knowledge about the population becomes available. Thus, the results of this Bayesian linear regression reduces to the ordinary linear regression, a commonly used statistical method. This model explains the relationship between county-level perceived risk, objective risks, and community resilience. The model is:

$$Y \sim N(\beta_0 + \sum_{i=1}^{N} \beta_i . X_i , \sigma^2) , \quad \text{Equation 6}$$

where $Y$ represents the county-level hurricane-related risk perceptions, $\beta_i$ are coefficients, $X_i$ are parameter values and $\sigma^2$ is the variance and is known. It is assumed that $\varepsilon_i \sim iid N(0, \sigma^2)$, $cov(x\ e) = 0$, $Var(y) = \sigma^2 I_n$. Under these assumptions, $E(y_i|\beta, X) = \beta_1 x_{1i} + \beta_2 x_{2i} + \cdots + \beta_k x_{ki}$ and $y|\beta, \sigma^2, X \sim N_n(X\beta, +\sigma^2 I)$.

Thus, the vector of unknown parameters to be estimates is $\theta = (\beta_1, \ldots, \beta_k, \sigma^2)$. Bayesian analysis requires specifying prior distributions for all model parameters, including the regression coefficients $\beta_i$. This study specifies Gaussian non-informative priors for each regression coefficient,

$$g(\beta, \sigma^2) \propto 1/\sigma^2 \qquad \text{Equation 7}$$

These priors are normally distributed and centered at zero with fixed variance, $\beta_i \sim N(0, 100^2)$. A large value of standard deviation is used to make the prior non-informative and proper, and reduce likely bias toward the mean (Gelman et al., 2004). This study uses a noninformative prior distribution as a convenient assumption for the purposes of explaining the relationship between risk and resilience. Other informative priors could be used in the next iterations of this research.

Bayesian inference is based on the posterior distribution for the parameters given the data. Thus, it combines the prior distribution with the likelihood. This study uses simulation to sample from the posterior distribution of the parameters. A Markov chain Monte Carlo (MCMC) method, specifically a Gibbs sampler is used to sample successively from the parameters' conditional posterior distribution (Gelman et al., 2004). The Gibbs sampler is a statistically efficient algorithm for posterior sampling.

To facilitate computation, both the response and predictor variables are standardized and centered. Model checking is conducted in four ways: (1) posterior predictive checking, with multiple choice of discrepancy measures to compare replicates of the dataset with the sample; (2) Bayes p-values, which measures the probability of getting more extreme residual sum of squares in replicated data then in the sample; (3) Residual plots to explore whether residuals are randomly distributed around zero, with constant variance, and (4) model comparison using Bayes factor. The model selection and validation techniques and results are available upon request. All computations are carried out in R using JAGS package (Just Another Gibbs Sampler).

# RESULTS

We present the results in two parts. The first section describes the results of the spatially transformed county-level risk perceptions. In the second section, we show the results of the regression model that tested for relationships between contextual hurricane risks and community resilience, on one hand, and county-level risk perceptions, on the other.

**County-level risk perceptions**

In this paper, we utilize a fairly new way of charactering risk perceptions at an aggregate level. The same spatial smoothing technique is used to aggregate individual-level vulnerabilities to the county level among Midwestern farmers (Gardezi and Arbuckle 2017). Figure 3 illustrates the county-level risk perceptions obtained after spatial smoothing average perceptions in each county. The map depicts regional clusters of risk perceptions. The phenomenon of spatial clustering depicted graphically in Figure 3 is confirmed statistically using Moran's I (Figure 4). We find that the computed value of Moran's I = 0.64 and the p-value is <0.001. Thus, neighboring counties have similar risk perception values (pattern is clustered). In other words, spatial autocorrelation is confirmed in this dataset. Moreover, a Monte-Carlo estimate of the p-value is calculated to ensure consistency in results.

The map illustrates that perceptions of hurricane-related risks are geographically heterogeneous, with heightened perceptions concentrating in central part of the Gulf coast. This area stretches from southeast Texas to west Florida. The overall pattern matches with that of peak heights of storm surge along the Gulf Coast since 1880 (Needham and Keim 2012). Storm surges with enormous peak heights of the past 100 years are clustered around the central part of the Gulf Coast. This suggests that past storm surge may have significant influence on shaping community-level perceptions of hurricane-related risks.

**Figure 3. County Level Risk Perception Scores**

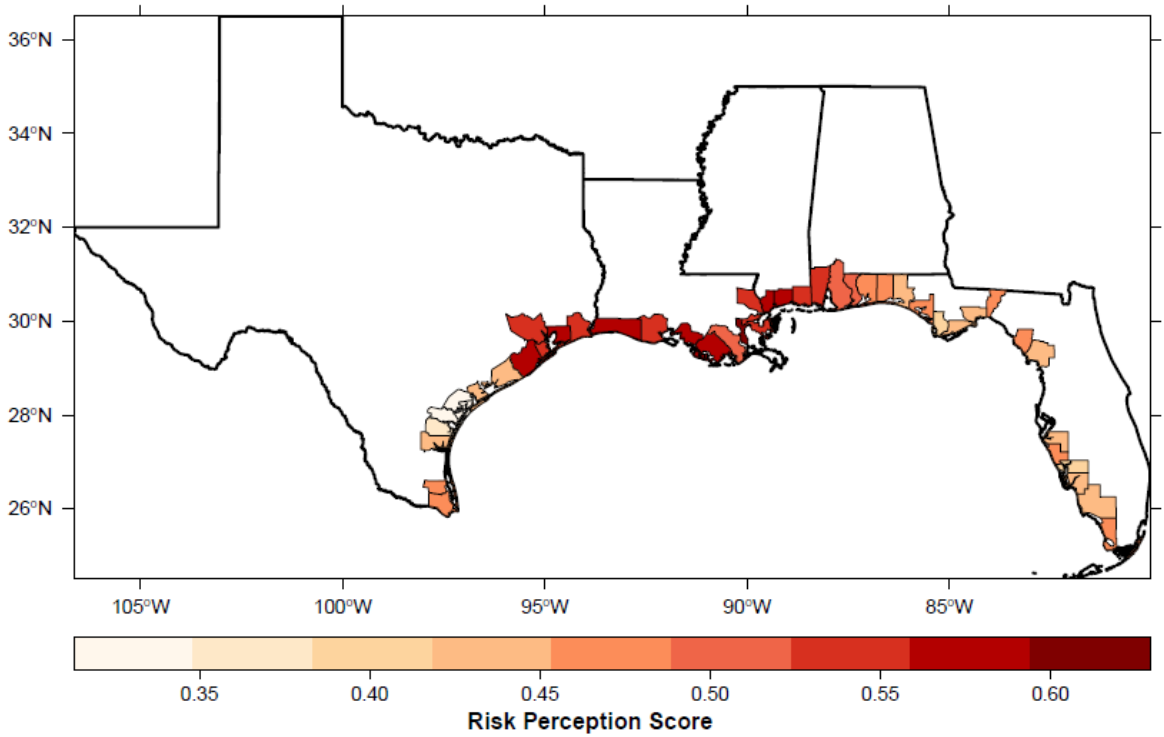

**Figure 4. Moran Plot, Moran I = .64**

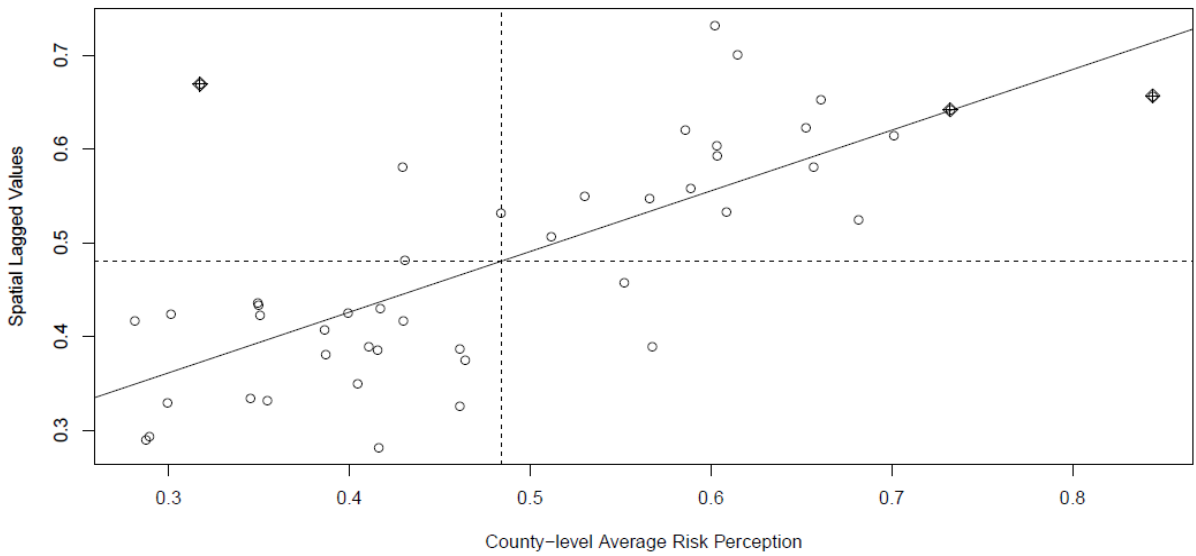

**Correlations among all variables**

The correlation graph (Figure 5) below highlights the strength of the relationship between hurricane risk perception and each community resilience indicator. Blue and red colored squares represent positive and negative correlation, respectively. The size of the square illustrates magnitude of the bi-variate correlation. As expected, county-level risk perception is positively correlated with all objective measures. The strongest correlation is between maximum wind speed from the last hurricane landfall and the county-level hurricane risk perceptions. Peak height of storm surge from the last hurricane landfall is also strongly correlated with the variable of interest. There are significant positive correlations between risk perception and indicators of social and economic resilience. Interestingly, land exposure to storm surge is positively correlated with environmental resilience, and land exposure to major hurricane wind zones is positively associated with infrastructural resilience. These correlations suggest that historical exposure to hurricane-related risks may exert significant influence on communities' efforts to increase their infrastructure and environmental resilience. More studies are needed to further reveal the relationship between exposure to hurricane hazards and each community resilience indicator.

**Regression results**

Table 5 summarizes the posterior means and 95% credible sets for the linear regression model along with their Bayes p-value and Deviance Information Criteria (DIC). The posterior means of the model results show that community resilience can influence risk perception at the county level. There is a positive relationship between social resilience, economic resilience, infrastructural resilience, community capital resilience, environmental resilience, respectively, and county level risk perceptions. However, institutional resilience and risk perception have an inverse relationship. There is also a positive relationship between the indicators of objective hurricane risks and county-level hurricane risk perceptions. For example, the posterior means of objective risks do not fall below the zero.

### Figure 5. Correlations among all variables

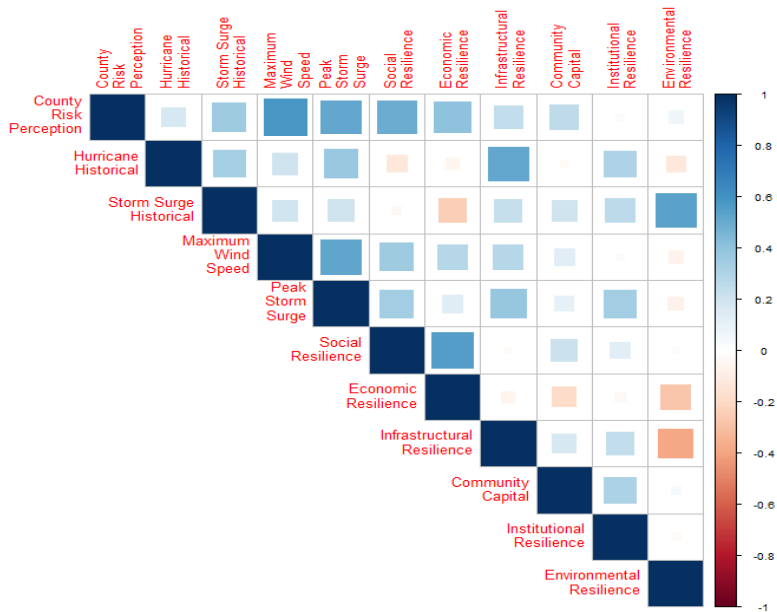

### Table 5. Summary of Bayesian linear regression model coefficients: Posterior means are reported (N = 46)

|  | Posterior Means | N=46 Lower 95% | Upper 95% |
|---|---|---|---|
| Intercept | 0.28 | 0.06 | 0.51 |
| **Community Resilience** | | | |
|     Social Resilience | 0.07 | -0.14 | 0.27 |
|     Economic Resilience | 0.33 | 0.11 | 0.56 |
|     Infrastructure Resilience | 0.04 | -0.15 | 0.22 |
|     Community Capital Resilience | 0.19 | 0.03 | 0.35 |
|     Institutional Resilience | -0.23 | -0.42 | -0.03 |
|     Environmental Resilience | 0.02 | -0.15 | 0.19 |
| **Objective Risks** | | | |
|     Maximum Wind Speed | 0.05 | -0.03 | 0.12 |
|     Peak Storm Surge | 0.08 | 0.01 | 0.15 |
|     Land Exposure to Hurricane Wind | 0.00 | -0.04 | 0.05 |
|     Land Exposure to Storm Surge | 0.09 | 0.01 | 0.17 |
| Bayes P-value | 0.45 | | |
| Deviance Information Criteria | 92.52 | | |

[1] Posterior means are reported with their 95% credible sets

Another intuitive way of illustrating these results is to examine the significance of all variables in the model. Figure 6 shows that the 95% credible sets for *Peak Height of Storm Surge associated with the most recent hurricane landfall*, *Exposure to Storm Surge*, *Institutional Resilience, Economic Resilience* and *Community Capital Resilience,* do not include zero, and thus, these variables are statistically significant. Therefore, exposure to high storm surge flooding and experience with higher storm surge in the recent past both heighten perceptions of hurricane-related risks at the county level. The results suggest that flooding risks generated by storm surge and topographic vulnerability can be salient for shaping community's perceptions of hurricane-related risk factors. Hurricanes often generate two types of hazards: excessive water and strong wind. Coastal residents may relate hurricane risks more with storm surge-induced flooding than wind as indicated by our results. Powerful hurricane winds can be destructive, incurring a great amount of damages. The policy implication is that the destructive power of both water and wind associated with hurricanes should be emphasized when local governments communicate hurricane threat with the public so that proper preventive measures such as purchasing flooding and wind insurance can be considered by residents.

Similarly, higher economic resilience and greater community capital resilience can lead to higher level of hurricane risk perceptions. These two results indicate that communities with more economic resources and social capital tend to perceive greater threat of hurricanes. To interpret the effects of economic resilience, environmental risks may be "luxury" that only well-off communities can afford to pay attention to. This argument is built upon Maslow's theory about human motivation (Maslow 1943). Specifically, humans tend to satisfy physiological needs such as food, shelter, safety before turning attention to other needs such as self-actualization and environmental quality. Community capital resilience estimated by Cutter, Ash, and Emrich (2014) represents a community's coherence when encountering environmental disasters. Unlike economic capital, this measure exhibits a community's social capital, "the propensity for a community to call on the good will of local citizens to assist their neighbors and fellow citizens" (Cutter Ash, and Emrich 2014; 68). Communities with higher level of unity and cohesion tend to have higher level of awareness of potential environmental risks because of community members' propensity and willingness to disseminate information they deem to be crucial to fellow citizens' well-being.

## Figure 6. Posterior Significance Levels

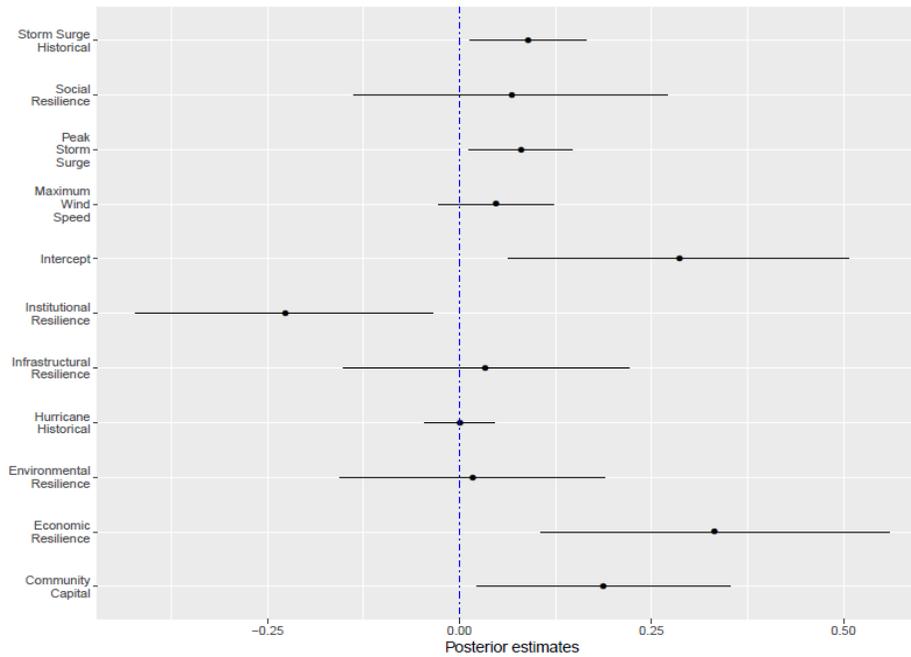

    The policy implication is that counties with less economic and social capitals need to direct efforts on educating the public about scientific assessments of hurricanes risks. The local government should consider utilizing a variety of venues such as social media, education campaigns, and public meetings to disseminate among these counties up-to-date scientific information regarding hurricane risks. Further, they need to utilize policy incentives to motivate individuals to build adaptive capacity to address environmental risks.

    Interestingly, higher level of institutional resilience is found to be correlated with lower levels of county level risk perceptions. This result can be interpreted as follows: institutional resilience can convey an immediate sense of security to the community. Public trust on the local government's capacity to cope with risks and hazards can build up under a healthy and resilient institutional environment. This can reduces public anxiety. For instance, communities that have witnessed intensive institutional efforts to mitigate natural disasters after the aftermath of an extreme event may experience a sense of relief, which nonetheless contributes to a lower state of risk perceptions.

## CONCLUSION

Utilizing the 2012 Gulf Coast Climate Change Survey merged with historical hurricane data and community resilience indicators, we first apply spatial statistical models to construct county-level risk perception indicators based on survey responses and then employ Bayesian regression to reveal the relationship between contextual hurricane risks and community resilience, on one hand, and county-level perceptions of hurricane risks, on the other. Our study presents an initial effort to examine perceptions of hurricane risks at the county level. Results of this study can be directly applicable in the policy-making domain as many hazard mitigation plans and policies are designed and implemented at the county level.

Specifically, we have made several important findings. First, among all measures of contextual hurricane risks, peak height of storm surge associated with the last hurricane landfall and land area exposed to historical storm surge positively affect county-level risk perceptions. Both measures are associated with one aspect of hurricane risks: water. It is known that hurricanes generate two types of hazards including excessive water and strong wind. The Saffir-Simpson scale, which is often used to quantify categories of hurricanes, is solely based on wind speed. However, as our results demonstrate, the damaging power of wind tends to be underestimated by people compared to storm surge when perceiving hurricane risks. The memory of storm surge rather than wind speed is likely to be invoked when being confronted with the question about hurricane-related risks. Given that hurricanes can incur damages through both water and wind, the local government should communicate hurricane risks to a full extent with the public. The two aspects of hurricane risks differ in their manifestations, which call upon different coping mechanisms. For instance, purchasing flood insurance can relieve excessive economic damages caused by storm surge while wind insurance is designed to recover from damages rendered by winds. In the meanwhile, heightened risk perception is a precondition for corresponding actions. Only when the awareness of wind risks associated with hurricanes is raised among coastal residents can proper mitigation measures be seriously considered and adopted. Local governments along the coast should utilize various venues to disseminate up-to-date scientific information among residents. For instance, intense education campaigns may manage to raise the awareness of hurricane risks.

Second, three out of the six community resilience indicators are found be significantly related to county-level perceptions of hurricane risks. Namely, communities with higher levels of economic resilience and community capital are more likely to perceive hurricane risks. This result point to the urgency to concentrate educational efforts and resources to raise the awareness of hurricane risks on communities with less economic and social capital. Over the long term, the most effective way would be to improve the levels of economic resilience and community capital in counties where those factors are low. Interestingly, institutional resilience is also a significant factor, though having negative impact on perceptions of hurricane risks. Our interpretation is that institutional resilience conveys a sense of security to the public who are more likely to trust the local government's competence to address hurricanes risks and thus less likely to perceive these risks. This also indicates the complexity of risk perceptions, including both risk appraisal and adaptive capacity appraisal. Future studies should consider the multi-facets of risk perceptions on the aggregate level.

It should be noted that there are a couple of limitations to the present study. First, hurricanes can bring triple threats to communities including wind, storm surge, and heavy rainfall. This study only considers the first two threats given that the study area is coastal with wind and storm surge being the dominant risks. To fully capture how hurricane risks affect community risk perceptions, future studies should consider inland communities since heavy rainfall from hurricanes can also lead to flooding, as manifested in Houston during Hurricane Harvey (2017). Second, the number of survey responses varies widely from county to county, which may affect the robustness of our statistical estimates. We have attempted to address this caveat using a simple random sampling approximation. However, future research can develop more robust methods of attending to unequal sample sizes for each county.

Last but not least, the research agenda on the dynamic relationships among contextual risks, community resilience, and risk perceptions is far from being complete. We propose a theoretical framework here (Figure 7). As the arrows suggest in Figure 7, because contextual risks are exogenous factors, they only impose impacts on the other two. Whereas the proposed relationship between community resilience and perception is not unidirectional. More empirical studies are needed to further examine our proposed relationships among these three components.

Specifically, first, more empirical studies using our approach are needed to examine how other environmental risks such as drought, heat wave, and inland flooding in different regions with community resilience impact risk perceptions. Second, environmental perceptions can be measured on three dimensions including: perceptions of past events, current conditions, and future probabilities, respectively. Each measure is meant to capture the subjective assessment of events in different temporal contexts. It is of interest to find out how the contextual risks in conjunction with community resilience impact the three aspects of perceptions. Third, although previous research points out the importance of including environmental risk perceptions in the construction of community resilience, no studies to date has taken this theoretical suggestion to an empirical level. Future studies need to incorporate the cognitive aspect of resilience such as risk perceptions and perceived adaptive capacity to deliver a more complete picture of community resilience.

**Figure 7. Proposed Theoretical Framework for Future Studies**

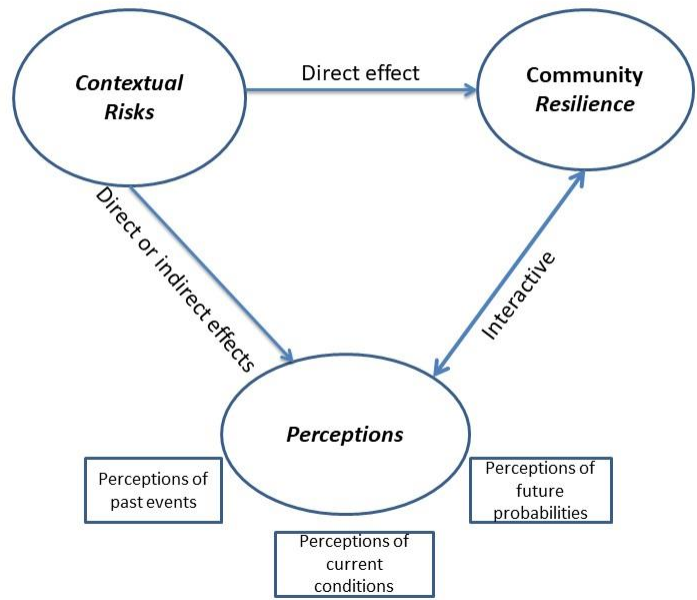

**Acknowledgements**

We would like to thank Kirby Goidel, LaDonn Swann, Tracy Sempier, and Melissa Schneider for their support in designing and implementing the 2012 Gulf Coast Climate Change Survey and the two reviewers for their suggestions and edits on this manuscript. The survey research included in the analysis was supported by the U.S. Department of Commerce's National Oceanic and Atmospheric Administration's Gulf of Mexico Coastal Storm Program under NOAA Award NA10OAR4170078, Texas Sea Grant, Louisiana Sea Grant, Florida Sea Grant, and Mississippi-Alabama Sea Grant Consortium. The views expressed herein do necessarily reflect the views of any of these organizations. Neither the organizations nor the individuals named above bear any responsibility for any remaining errors. Wanyun Shao is supported by National Academy of Sciences, Engineering, and Medicine Gulf Research Program Early-Career Research Fellowship. Siyuan Xian is supported by National Science Foundation (NSF) grant: EAR-1520683